\documentclass[graybox]{svmult}

\usepackage{mathptmx}       
\usepackage{helvet}         
\usepackage{courier}        
\usepackage{type1cm}        
\usepackage{makeidx}         
\usepackage{graphicx}        
\usepackage{multicol}        
\usepackage[bottom]{footmisc}

\usepackage{bm}
\usepackage{xcolor}
\usepackage{hyperref}
\usepackage{makeidx}
\usepackage{amsmath}
\usepackage{graphicx}
\usepackage{dcolumn}
\usepackage{amsfonts}
\usepackage{amssymb}
\usepackage[latin9]{inputenc}
\usepackage{color}

\makeindex             

\begin{document}

\title*{Finite Size Effects in Topological Quantum Phase Transitions}
\author{Mucio A. Continentino, Sabrina Rufo and Griffith M. Rufo}

\institute{Centro Brasileiro de Pesquisa F\'{\i}sicas, Rua Dr. Xavier Sigaud, 150, Rio de Janeiro, RJ, 22290-180,  Brazil \email{mucio@cbpf.br}
}

\maketitle
\abstract{The interest in the topological properties of materials brings  into question the problem of topological phase transitions. As a control parameter is varied, one may drive a system through phases with different topological properties. What is the nature of these transitions and how  can we characterize them?   The usual Landau approach, with the concept of an order parameter that is finite in a symmetry broken phase is not useful in this context. Topological transitions do not imply a change of symmetry and there is no obvious order parameter.  A crucial observation is  that they are associated with a diverging length that allows a scaling approach and to introduce critical exponents which define their universality classes. At zero temperature the critical exponents obey a quantum hyperscaling relation. We study finite size effects at topological transitions and show they exhibit universal behavior due to scaling. We discuss the possibility that they become  discontinuous as a consequence of these effects and point out the relevance of our study for real systems.}

\section{Topological phase transitions}
\label{sec:1}
Topology studies the stability of forms, shapes under different operations. These may occur in abstract spaces as in momentum space reciprocal to crystalline structures~\cite{alicea,kane,shen}. If certain symmetries are present, they give rise to invariants that are robust under different operations. In many cases, in condensed matter systems, these topological invariants are directly related to physical observables~\cite{alicea,kane,shen}. The existence of non-trivial topological phases derives from their symmetry properties, but may occur only for restricted regions of the parameter space characterizing the system. As a consequence, if these parameters are changed, the system may transit from one non-trivial topological phase to another or even to a trivial topological phase. Here we will be interested in  topological transitions that occur at zero temperature ($T=0$)~\cite{physicab}, as a physical parameter like the chemical potential is varied. The critical fluctuations in this case are purely quantum mechanical and the topological transition is a quantum phase transition~\cite{book}.
These phase transitions that are of great interest nowadays differ~\cite{koster}, but also share many features with conventional ones. A significant difference is the lack of an order parameter since in general there is no symmetry breaking  at a topological transition.  The use of a topological invariant as an order parameter is not a valid option as it changes abruptly. This  may wrongly suggest that the phase transition is discontinuous and does not fully develops. The main consequence of the absence of an order parameter is that a Landau expansion~\cite{landau} of the ground state energy in terms of a small quantity near the transition is not possible.
\begin{svgraybox}
The most important feature that characterizes a topological transition as a genuine critical phenomenon  is the existence of a characteristic length  $\xi$ that diverges at this transition. If $g$ is a control parameter, such that, the transition occurs at $g=0$, we can write
\begin{equation}
\xi = \xi_0|g|^{-\nu},
\end{equation}
where we will refer to $\nu$ as the {\it correlation length exponent} and $\xi_0$ is a natural length of the system, as the lattice spacing.
\end{svgraybox}
The identification of this characteristic length is guided by a unique attribute of non-trivial topological phases, namely,  the existence of surface states that decay as they penetrate the bulk of the material~\cite{kitaev}. This {\it penetration length}  diverges at the topological transition and can be identified as  the characteristic length scale associated with  this critical phenomenon~\cite{physicab,RG,cristiane,nandini,chen,chen2}. 

\begin{svgraybox}
The existence of this  diverging length  allows to develop a scaling theory for topological transitions~\cite{book}.
The singular part of the temperature  dependent free energy as a function of the distance $g$ to the transition  can be written as~\cite{book},
\begin{equation}
\label{freesca}
f_s \propto |g|^{\nu(d+z)}F\left[\frac{T}{|g|^{\nu z}}\right].
\end{equation}
If hyperscaling holds, the quantum hyperscaling relation implies
\begin{equation}
\label{hyper}
2-\alpha=\nu(d+z),
\end{equation}
where we  introduced two new quantum critical exponents, $\alpha$ and $z$. Since the scaling function $F[0] =$ constant, the former characterizes the singular behavior of the ground state energy density~\cite{mucio}. The latter is the dynamic critical exponent and $d$ is the dimension of the system.
\end{svgraybox}

The dynamic critical exponent $z$  plays a fundamental role in quantum critical phenomena~\cite{book}. Here, it is  defined by the form of the  the dispersion relation of the excitations at the QCP, $g=0$, i.e., $\omega(g=0) \propto k^z$. In general for isotropic systems close to the topological transition, the spectrum of excitations  can be written as, $\omega =\sqrt{|g|^{2 \nu z} + k^{2z}}$~\cite{griffith}. The {\it wavevector}  $k$ is that for which the gap $\Delta = |g|^{\nu z}$ closes at the transition.   In the cases of interest here the dynamic exponent $z$ takes the Lorentz invariant value $z=1$, as a consequence of the Dirac-like nature of the dispersion relation {\it at} the transition~\cite{griffith}. 

It is important to mention that the quantum hyperscaling relation, Eq.~\ref{hyper}, that relates the quantum critical exponents to the dimension of the system can be violated in several ways~\cite{book}. For example, when the critical exponent $\alpha$ determined by this relation becomes negative. For the systems studied here with $z=1$ and $\nu=1$, as obtained below, this  occurs for $d>1$. In this case there may be analytic contributions to the free energy, like $f \propto |g|^2$ that for  $\alpha < 0$ will vanish more slowly close to the QCP than the scaling contribution~\cite{analytic}.  This implies that the exponent $\alpha$ remains fixed at $\alpha=0$ for all $d \ge 1$. For $d=1$, the marginal dimension, there may be also  logarithmic corrections for the ground state energy (see below).
Hyperscaling may also breakdown if the dispersion relation of the system is highly anisotropic, such that, the correlation length exponent is not uniquely defined but depends on a given direction~\cite{anisotropic}.

Notice that in conventional quantum phase transitions the algebraic decay of correlations of the order parameter at the QCP requires introducing a critical exponent $\eta$~\cite{book}. This is related to the exponent $\beta$ of the order parameter through another hyperscaling relation $2  \beta=\nu(d+z-2+\eta)$~\cite{book}. The exponents $\eta$ and $\beta$ play no role in the characterization of topological quantum phase transitions as discussed here.

In the next Sections, we study two models exhibiting  topological transitions and determine their universality classes, essentially the critical exponents $\nu$, $z$ and $\alpha$.   We start with the  one-dimensional ($1d$) Su-Schrieffer-Heeger  (SSH)~\cite{shen} model for a dimerized tight-binding chain, which is one of the simplest model to exhibit a quantum topological phase transition. We also consider the two-dimensional   ($2d$) Bernevig-Hugues-Zhang model~\cite{shen} and obtain  the correlation length exponents $\nu$ for both  models. Finally, we discuss a $3d$ model of a topological insulator and the possible occurrence of a discontinuous transition in this system.

\section{The Su-Schrieffer-Heeger model}

The Su-Schrieffer-Heeger (SSH) model~\cite{shen} has been proposed to study the electronic properties of the polymer composed of repeating units of polyacetylene organic molecules ${ (C_2H_2}) _n$. The  Hamiltonian in real space can be written as

\begin{eqnarray}\label{SSHrealspace}
\mathcal{H}=\sum_{n} \psi_n^\dagger A \psi_n +  \psi_n^\dagger B \psi_{n-1} + \psi_n^\dagger B^\dagger \psi_{n+1},
\end{eqnarray}
where $\psi_n= (\psi^a_n,\psi^b_n)^T$ is the  wave function vector of a unit cell $n$ with wave function components $\psi^a$ and $\psi^b$ from $a$ and $b$ sublattices, respectively. The intra and inter cell hoppings are given by $2 \times 2$ matrices  $(A)_{i,j} = t^{*}_1 \delta_{i, j-1} + t_1 \delta_{i-1, j} $ and $(B)_{i,j} = t^{*}_2 \delta_{i,j-1} $, respectively, where $t_1$ and $t_2$ are real numbers that represent the intra and inter cell hopping terms.
After a Fourier transformation of the Hamiltonian, Eq~\ref{SSHrealspace}, we get
\begin{eqnarray}\label{SSHkspace}
\mathcal{H}=\sum_{k} \psi_k^\dagger H(k) \psi_k,
\end{eqnarray}
such that,  $\psi_k =(\psi^a_k,\psi^b_k)^T $ and  $(H(k))_{i,j} = t(k) \delta_{i,j-1} +t^{*}(k) \delta_{i-1,j}$ with $t(k) = t_1+t_2 e^{ika}$.
A diagonalization process allows to obtain the energies of the electronic states of the model as

\begin{equation}\label{energyk}
  E(k) = \pm |t(k)| = \pm \sqrt{t^2_1+t^2_2+2t_1t_2\cos k},
\end{equation}
where the lattice spacing was taken equal to unity. Notice that, for $|t_1| \neq |t_2|$, this energy dispersion presents a gap around  zero energy. Therefore,  if the Fermi level $\mu$ is taken at zero energy, the ground state describes an insulating phase. On the other hand, this model undergoes a topological phase transition at the quantum critical point, $g=t_1-t_2=0$, with a gap closing at  $k=\pi$.

The insulating phase that arises when $|t_1|>|t_2|$ is a trivial topological phase, since the topological invariant winding number $W$ is equal to zero. For $|t_1|<|t_2|$, the insulating phase is topologically non-trivial with winding number equal to one. In the topological non-trivial phase, there are edge states with zero energy, $ (\psi^a_n(E=0),\psi^b_n(E=0))$,  that are protected by the topology of the Bloch bulk electronic states.

Solving recursively for the zero-energy eigenstates of the Hamiltonian Eq~\ref{SSHrealspace}, we find for the ratio of the wave functions at sites $n$ and  $1$ at the edge of the $a$ sub-lattice,
\begin{equation}\label{psia}
\delta \psi^a_n= \frac{\psi^a_n(E=0)}{\psi^a_1(E=0)}=  \left( -\frac{t_1}{t_2} \right)^{n-1}.
\end{equation}%
These edge states are mostly located at the edges of the chain, more precisely in the unit cells 1 and N of the SSH model. Their existence  is guaranteed by the condition $E=0$ in Eq.~\ref{energyk} that leads to $t(\tilde{k}_0)=0$ or $e^{i\tilde{k}_0}=-(t_1/t_2)$.  Notice that in the case of edges states with zero energy, $\tilde{k}_0$ is a complex number. Substituting,  $(t_1/t_2) = -e^{i\tilde{k}_0}$ in Eq.~\ref{psia}, the ratio of  wave functions for the $a$ sub-lattice can be written as
\begin{equation}\label{exp1}
\delta   \psi^a_n= e^{i\tilde{k}_0(n-1)}.
\end{equation}
The value of $\tilde{k}_0$ as a function of the distance from the critical point can be obtained from the following equation
\begin{equation}\label{ExpEnery}
E(k) \sim \sqrt{ g^2 + t_1 t_2 k^2},
\end{equation}
which is a series expansion of Eq.~\ref{energyk} near the PCQ. We have introduced $g=t_1-t_2$ to represent the distance from this PCQ. For an edge state $E(\tilde{k}_0)=0$ and therefore  Eq.~\ref{ExpEnery} yields $\tilde{k}_0 = i (g/\sqrt{t_1 t_2 })$.

Finally, substituting $\tilde{k}_0$ in Eq~\ref{exp1}, we obtain for the wave functions ratio
\begin{equation}\label{expfinal}
\delta \psi^a_n = e^{-(n-1)/\xi},
\end{equation}
where $\xi=\sqrt{t_1 t_2 } |g|^{-1}$. The normalized wave function decays exponentially with $n$ within the bulk with a penetration depth $\xi$  that diverges with critical exponent $\nu=1$. Notice that this result can also be obtained directly from Eq.~\ref{psia}.

\begin{figure}
  \centering
  \includegraphics[scale=0.6]{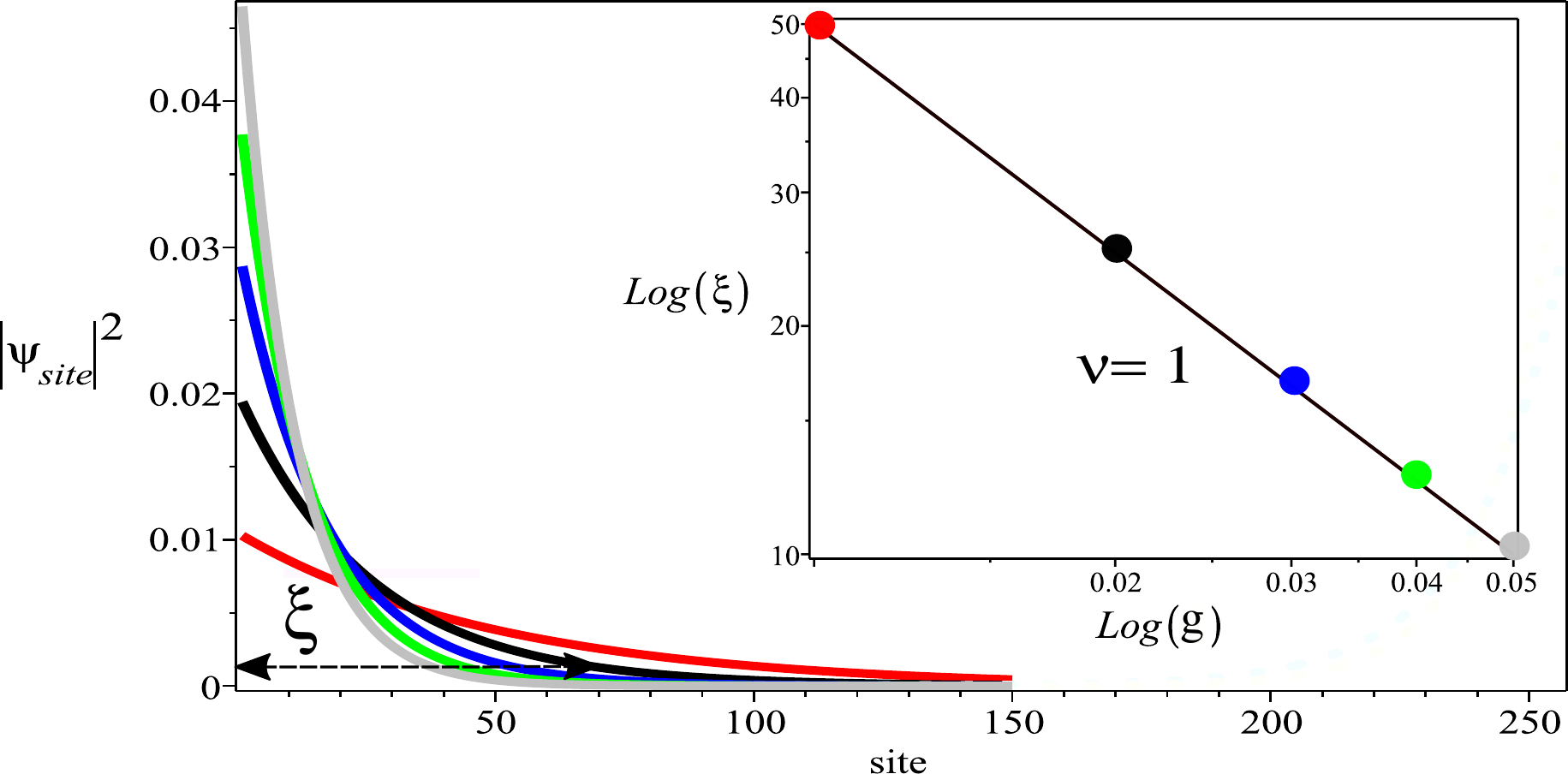}
\caption{The square of the wave function (Edge states) as a function of the sites. Solid lines are solutions for the $a$ sublattice.
The penetration depth $\xi$ is shown for $g=0.02$ (black curve). At $\xi$ the wave function satisfies the condition $\Psi_n(\xi) =\Psi_1/e $.
The inset shows the penetration depth versus $g$. Different colors represent different values of $g$ as depicted in the inset. The angular coefficient of the straight line is formally the critical exponent, $\nu=1$.}
\label{fig1}
\end{figure}

Fig.~\ref{fig1}  shows the square of the  wave functions ( $| \psi^a_n(E=0)|^{2}$ ) of the edge states  obtained numerically from Eq.~\ref{SSHrealspace}.  The solid lines in the figure are the solutions for $a$ sub-lattice as a function of the sites.
There are similar solutions for the $b$ sub-lattice (not shown) that in this case are localized near the last site.

We have defined the penetration depth $\xi$ as the distance, relative from the initial site, for which $\Psi_n(x_0+\xi) =\Psi_1(x_0)/e $. Now, by considering several values of $g$, varying from $g=0.01 $ to $0.05$, we obtain $\xi(g)$ and determine the critical exponent $\nu$, as shown in the inset of  Fig.~\ref{fig1}. We get,  $\nu=1$ in perfect agreement with the analytic result of Eq.~\ref{expfinal}, showing that the numerical method is very reliable. In the next Section, we use the numerical  approach  to obtain the critical exponent $\nu$ of the $2d$ BHZ model.

\section{The Bernevig-Hugues-Zhang (BHZ) model}

The first experimental observation of a $2d$ topological quantum phase transition was in a CdTe/HgTe/CdTe heterostructure. This consisted of a layer of HgTe sandwiched between CdTe yielding a semiconductor quantum well~\cite{konig}. At some critical thickness value of these quantum wells, the topological quantum phase transition takes place, from a conventional insulating phase to a quantum Hall effect phase with helical edge states protected by the non-trivial topology of the bulk. This topological quantum phase transition can be described by the BHZ model~\cite{BHZ} associated with the following Hamiltonian
\begin{eqnarray}\label{BHZkspace}
H(k_{x},k_{y})= \vec{\sigma}\cdot \vec{h}(k),
\end{eqnarray}
where $\vec{h}(k)$ takes values on the two-dimensional Brillouin zone ($k_{x}, k_{y}$) and   $\vec{\sigma}=\{\sigma_{x}, \sigma_{y}, \sigma_{z}\}$ are the Pauli matrices. Specifically, $h_{x}=t_{sp}\sin{k_{x}}$, $h_{y}=t_{sp}\sin{k_{y}}$ and $h_{z}=2t_{1}(\cos{k_{x}}+\cos{k_{y}})+t_{2}-4t_{1}$. In this Hamiltonian, the sub-lattice space represents the orbitals $s$ and $p$ for each atom. In order to describe the quantum wells in HgTe/CdTe layers, the simplified spinless BHZ model introduces the hopping terms $t_{sp}$ and $t_{1}$, as well as, a {\it mass} term $t_{2}$. The antisymmetric hybridization between the  orbitals of different parities, $s$ and $p$ has an amplitude given by $t_{sp}$, and the hopping between the same orbitals $s$ or $p$ of nearest neighbors atoms has an amplitude $t_{1}$.

A topological phase can be identified by some proper topological invariant. For the $2d$ BHZ model, we can consider the Chern number invariant $\mathcal{C}$~\cite{bernevighughes,hasan} obtained at the high-symmetry points $[k_{x}, k_{y}]=\{[0,0], [0,\pi], [\pi,0], [\pi, \pi]\}$. It predicts a non-trivial topological phase for the intervals $0<t_{2}<4t_{1}$ with $\mathcal{C}=1$ and $4t_{1}<t_{2}<8t_{1}$ with $\mathcal{C}=-1$. A trivial phase with $\mathcal{C}=0$ occurs for $t_{2}>8t_{1}$. The Chern number signs $\mathcal{C}=\pm1$ are related to edge states with propagation in opposite directions.

Here we are interested in determining numerically the correlation length critical exponent for the two-dimensional BHZ model. For this purpose,  we study the penetration of the edge states, which requires one of the dimensions of the lattice to be finite.  Since these edge states are indeed connected to the real terminations of the system, for a square lattice to keep one of the dimensions finite means to deform the lattice into a cylinder. The finite axis takes the direction of the main axis of the cylinder and the other dimension with periodic boundary conditions is represented by the body of the cylinder.

The correlation length critical exponent as before characterizes the decay of the edge states into the bulk close to the topological transition. Let us consider the edge states of the BHZ model in one dimension. One way to get one of the dimensions finite is to perform a Fourier transformation as
\begin{equation}\label{somakx}
{H}_{I,J}(k_{y})=\frac{1}{N_{x}}\sum_{k_{x}}e^{ik_{x}(m-{m^{\prime}})}{H}_{I,J}(k_{x},k_{y}),
\end{equation}
where $N_{x}$ is the number of sites along the finite $x$-axis and $\{I,J\}$ indexes run over the matrix elements of Eq.~\ref{BHZkspace}. The positions of the atoms along the finite $x$-axis are denoted by $m$ and vary from $0$ to $N_{x}$. For example, considering the element ${H}_{1,1}(k_{x},k_{y})=h_{z}$ we have
\begin{eqnarray}\label{somakxH11}
{H}_{1,1}(k_{y})&=&\frac{1}{N_{x}}\sum_{k_{x}}e^{ik_{x}(m-{m^{\prime}})}{H}_{1,1}(k_{x},k_{y}) \nonumber \\
&=&\frac{1}{N_{x}}\sum_{k_{x}}e^{ik_{x}(m-{m^{\prime}})}\left[ 2t_{1}\cos{k_{x}}+C \right] \nonumber \\
&=&\frac{1}{N_{x}}\sum_{k_{x}}e^{ik_{x}(m-{m^{\prime}})} \left[ t_{1}(e^{ik_{x}}+e^{-ik_{x}})+C \right] \nonumber \\
&=&\frac{1}{N_{x}}\sum_{k_{x}}\left[ t_{1}\left(e^{ik_{x}(m-{m^{\prime}}+1)}+e^{ik_{x}(m-{m^{\prime}}-1)}\right)+Ce^{ik_{x}(m-{m^{\prime}})}\right] \nonumber \\
&=&t_{1}[\delta_{m,m^{\prime}+1}+\delta_{m,m^{\prime}-1}]+C\delta_{m,m^{\prime}},
\end{eqnarray}
where $C=2t_{1}\cos{k_{y}}+(t_{2}-4t_{1})$ is independent of $k_{x}$ and the same procedure should be applied to all the other matrix elements.

The sum over $k_{x}$ allows  to work in  real space along the $x$-axis. Notice that  we chose the $k_{x}$ to be in the finite direction, but since we consider a square lattice the choice between $k_{x}$ or $k_{y}$ is irrelevant due the symmetry of the lattice. For the diagonal directions of the square lattice~\cite{Imura}, or for  more complex lattices this is not necessarily true.  For instance, for the honeycomb lattice, the choice of the finite axis along one or other direction means different edge arrangements~\cite{Imura2}.

Accordingly, after  Fourier transforming Eq.~\ref{somakx},  from momentum to real space along the $x$-axis, we have
\begin{equation}\label{BHZrealspacex}
H(k_{y})=  \left(
    \begin{array}{cc}
      H_{11} & H_{12} \\
      H_{21} & H_{22} \\
    \end{array}
  \right).
\end{equation}
Following the procedure of Eq.~\ref{somakxH11} yields $H_{11}=[2t_{1}\cos{k_{y}}+(t_{2}-4t_{1})]\delta_{m,m^{\prime}}+t_{1}[\delta_{m,m^{\prime}+1}+\delta_{m,m^{\prime}-1}]$ that stands for sub-lattice $a$ and $H_{22}=-H_{11}$ for sub-lattice $b$. Here, the sub-lattices indexes $a$ and $b$ represent the subspace of the orbitals $s$ and $p$, respectively. The matrix elements responsible for the mixing of the different orbitals or sub-lattices are given by $H_{12}=-it_{sp}\sin{k_{y}}\delta_{m,m^{\prime}}-\frac{it_{sp}}{2}[\delta_{m,m^{\prime}-1}-\delta_{m,m^{\prime}+1}]$ and $H_{21}=H_{12}^{\dagger}$. The $m$ index counts the unit cells or atoms along the finite x-axis and in the same way $m^{\prime}$ can be interpreted as a neighbor site in the real space Hamiltonian, Eq.~\ref{BHZrealspacex}. Besides, the order of each matrix element $H_{I,J}$ is increased to $N_{x}\times N_{x}$, which means that the order of the final matrix  becomes $2N_{x}\times 2N_{x}$.

\begin{figure}[t]
\sidecaption[t]
\includegraphics[scale=0.3]{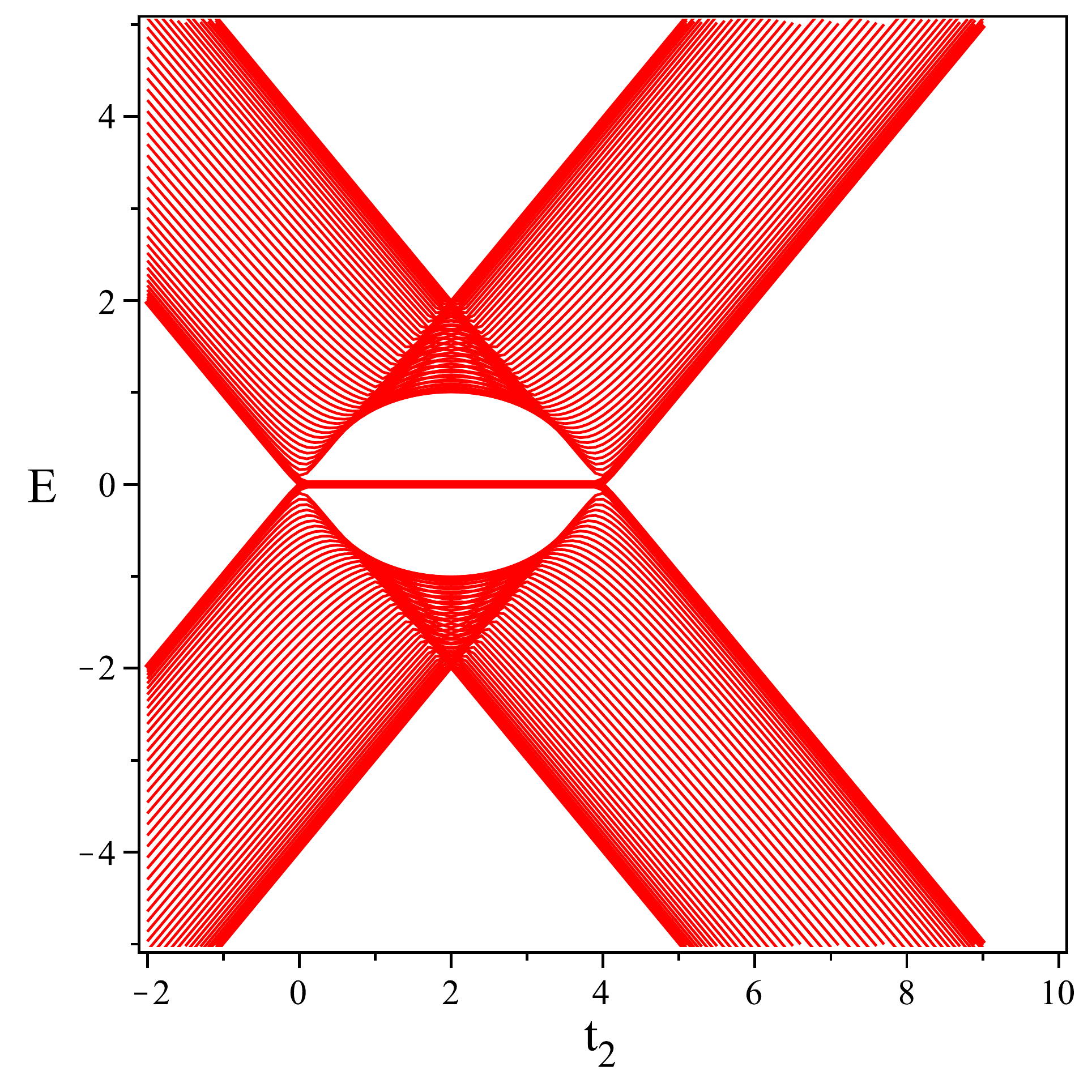}
\caption{(Color online) Energy dispersion of the BHZ model as a function of $t_{2}$, for a fixed $t_{1}=1$. The red lines are obtained for $k_{y}=0$ and for $N_{x}=50$ sites. Topological quantum phase transition takes place for $t_{2}=0$ and $t_{2}=4$. Along the line E=0, we highlighted the presence of the two edge states with thick lines.}
\label{fig2}
\end{figure}
\begin{figure}[t]
\sidecaption[t]
\includegraphics[scale=0.3]{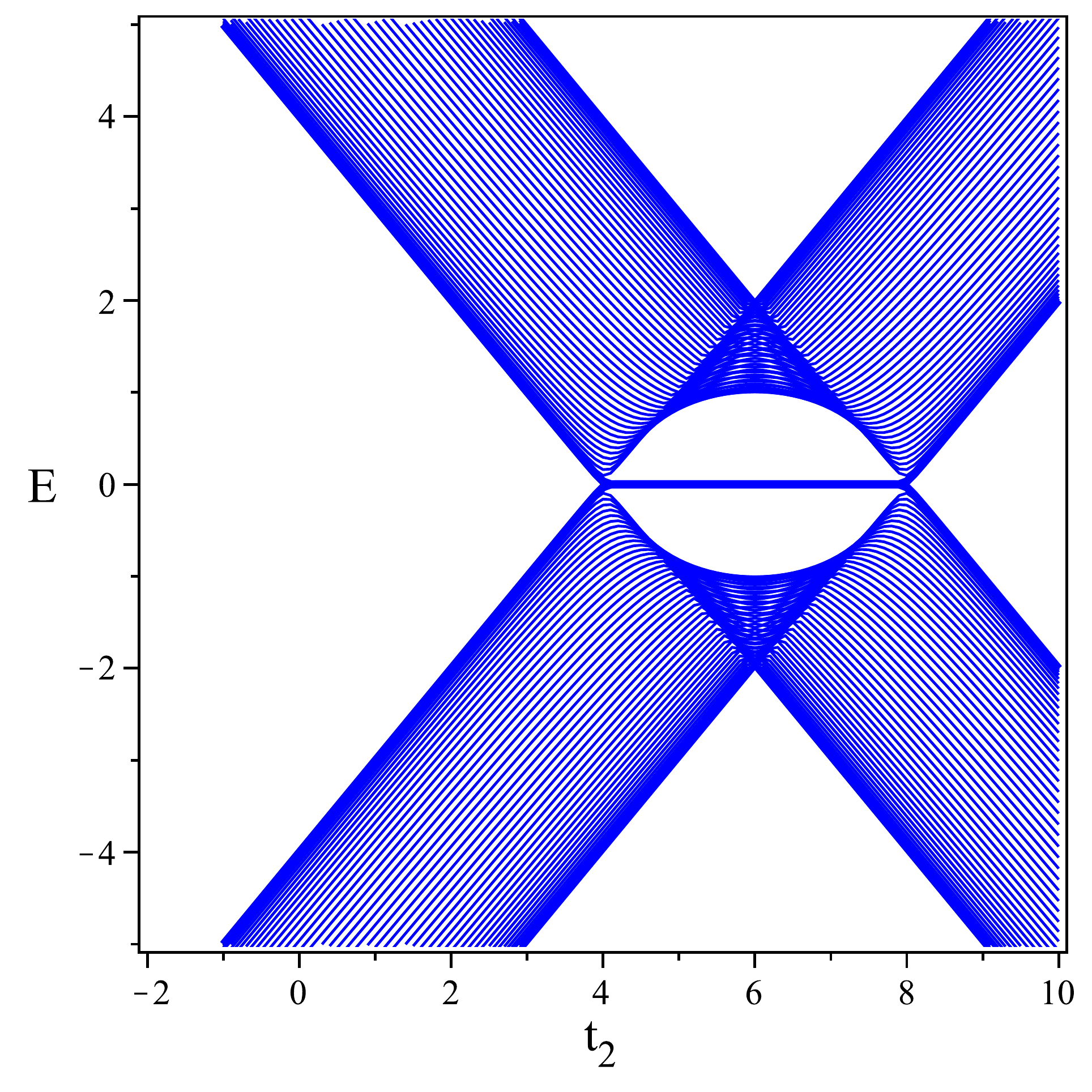}
\caption{(Color online) The same of Fig.~\ref{fig2}, but the blue lines are obtained for $k_{y}=\pi$. In this case, the topological quantum phase transition takes place for $t_{2}=4$ and $t_{2}=8$. Again, we highlighted the presence of the two edge states with thick lines along E=0.}
\label{fig3}
\end{figure}

For the purpose of obtaining the energy dispersion in real space, a numerical study of the $2d$ BHZ model was developed to diagonalize  the Hamiltonian, Eq.~\ref{BHZrealspacex}. We fix  the energy scale as $t_{1}=1$ and take $N_{x}=50$ sites. In  Figs.~\ref{fig2} and~\ref{fig3}, respectively, we present the energy $E$ as a function of the topological transition control parameter (mass)  $t_{2}$ at the high symmetry points,  $k_{y}=0$ and $k_{y}=\pi$. In the first case, $k_{y}=0$,  topological quantum phase transitions take place for $t_{2}=0$ and $t_{2}=4$. The thick lines in the figures show the presence of the edge states with zero energy. The same is observed for $k_{y}=\pi$, but the transition points are now given by $t_{2}=4$ and $t_{2}=8$.

\begin{figure}[t]
\centering
\includegraphics[scale=0.7]{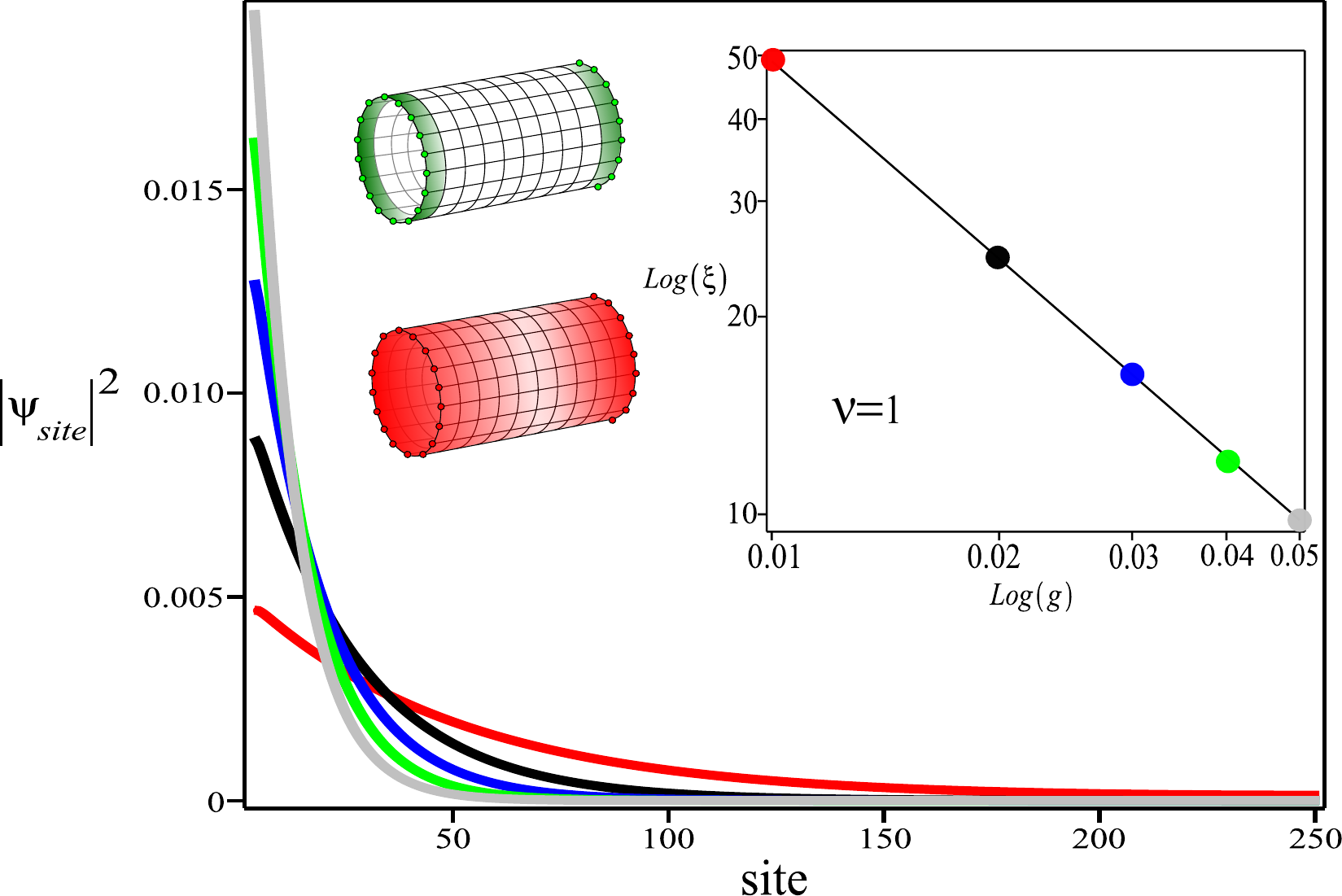}
\caption{The square of the wave function (edge states) as a function of sites ($N_{x}=500$) sites close to the quantum critical point, $t_{2}=0$. We show just half of the lattice ($N_{x}=250$) for one sub-lattice, since the behavior is the same on both sides for the two sub-lattices $a$ and $b$. The color scheme and the penetration depth follows that of Fig.~\ref{fig1}.
The inset shows the penetration depth versus $g$ and yields with high accuracy the value $\nu=1$ for the critical exponent of the penetration depth.
For all the QCPs as $g$ increases and the system moves away from these QCPs, the edge states become more localized at the ends.  The cylinders represent the finite BHZ lattice along the $x$-axis direction (main axis of the cylinder) with periodic boundary conditions in $y$-axis (body of the cylinder).
The top cylinder shows the penetration of the wave-function of the edge states for $g=0.04$ according to the color gradient (green). Similarly, the bottom cylinder shows the same (red), but for $g=0.01$.  In this case the wave function of the edge state penetrates almost the entire lattice.}
\label{fig4}
\end{figure}

For the study of the penetration of the edge states, we identify the eigenvectors responsible for the zero energy dispersions in Figs.~\ref{fig2} and~\ref{fig3}. For $N_{x}=500$,  we show in Fig.~\ref{fig4} the square of the wave function of the edge states in the vicinity of the critical points. Actually, just  one half of the lattice is presented, since the behavior is the same on both sides. In addition, the  results for sub-lattice $a$ and $b$ coincide. The edge states are obtained for distances to the critical point ranging from $g=0.01$ to $g=0.05$. As $g$ increases, the edge states become more localized at the edges of the lattice. The inset presents the characteristic length $\xi$ as a function of $g$ and the points are obtained from the numerical study of the model. From the linear fitting of these points, we can conclude with accuracy that the correlation length critical exponent  for the $2d$ BHZ model   is $\nu=1$. As mentioned before, in real space the lattice is a cylinder and   Fig.~\ref{fig4}  presents a pictorial view of the penetration of the edge states from the perspective of this cylinder. The top cylinder represents the case where the edge states penetration decays very fast. The color gradient follows the penetration intensity of the edge state. In the same way, the bottom cylinder shows a case where the edge state extends almost along the entire lattice. The color gradient here holds inside the cylinder body. These results reflect strictly the behavior obtained for all critical points $t_{2}=0$, $t_{2}=4$ and $t_{2}=8$.

\begin{figure}[t]
\centering
\includegraphics[scale=0.7]{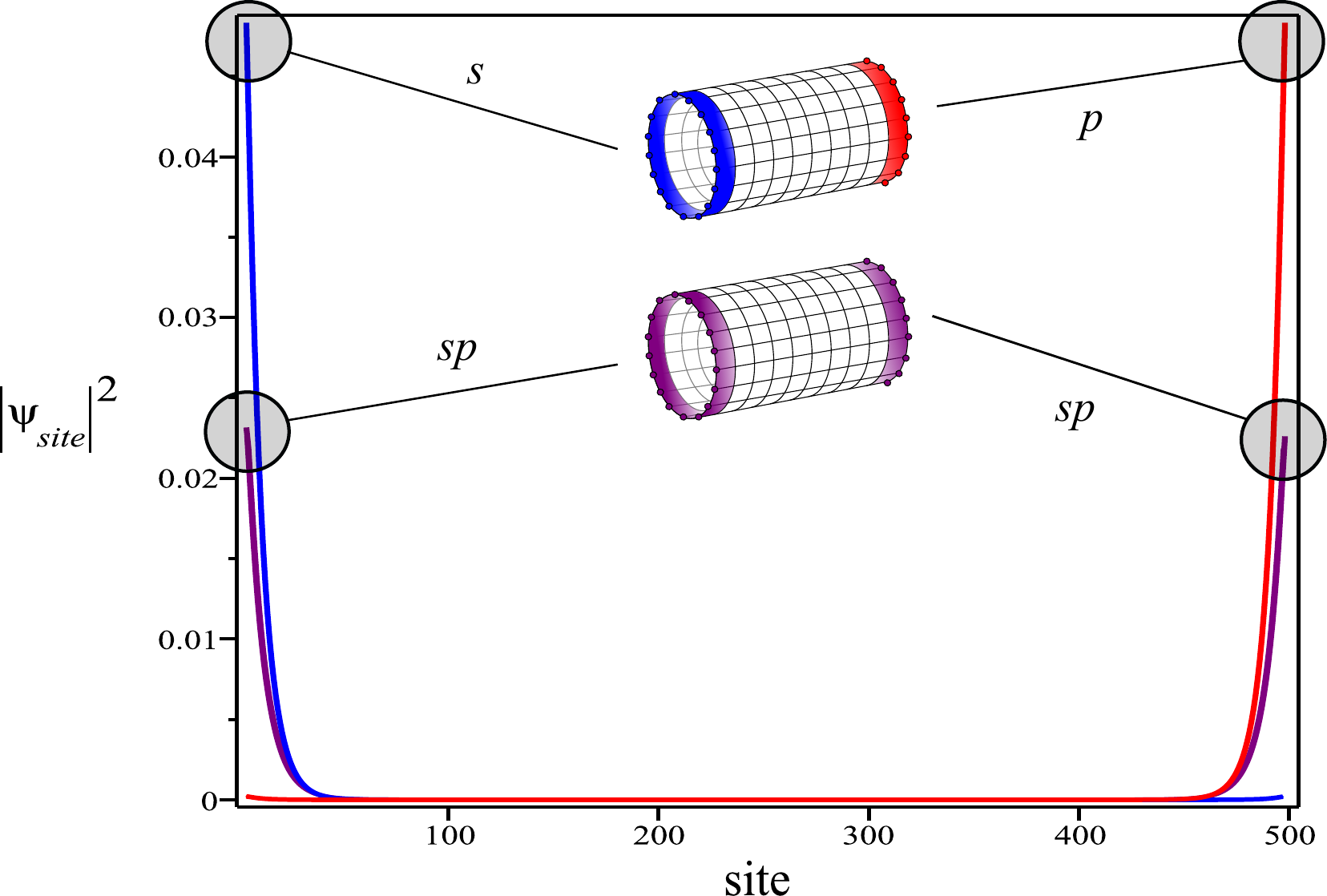}
\caption{(Color online)
The same of Fig.~\ref{fig4}, but exploring the effect of the proximity to criticality on the nature of the edge states. In the close vicinity of the topological quantum phase transition the edge states of the subspace of the orbitals $s$ and $p$, in purple color, coincide and exhibit the same amplitudes in both sides of the lattice. This is represented by the bottom cylinder filled with the same purple color in each edge. However, far away from the critical point ($ g \ge 0.068)$ the edge states on different sides begin to present mostly $s$ (left) or $p$ (right) character.   For instance, the wave function of the edge state nearly localized at the left end has a very strong $s$character. The opposite for the right end with a wave function with mostly $p$-character. This is indicated by the top cylinder where the blue and red colors represent the  $s$ and $p$ orbitals, respectively.}
\label{fig5}
\end{figure}

In the process of varying the distance to the quantum critical point, we notice that as the system moves away from the QCP, the behavior of the penetration length for the orbitals  $s$ and $p$ (sub-lattices $a$ and $b$) becomes distinct at the different edges.   Fig.~\ref{fig5} shows that for $g \geq 0.068$,  the wave function of the left edge state is nearly localized and has mostly $s$-character, while that of the right edge has mostly $p$-character. We also observe that the amplitude of the wave functions at the edges and consequently their  localization at these sites becomes larger as $g$ increases. The cylinders here indicate the correspondence between the edge states of the subspaces and the termination of the lattice for each case.
Finally, close to the QCP the wave functions of the edge states have a mixed character, as shown in Fig.~\ref{fig4}, due to their strong hybridization.

In summary our numerical study of the $2d$ BHZ model shows that the critical exponent for the penetration depth takes the value $\nu=1$, the same we have obtained for the $1d$ SSH and for a 1d $sp$-chain~\cite{book}. We have also pointed out a qualitative change in the nature of the edge states for the 2d BHZ model as the distance to the QCP of the topological transition changes.

\section{Finite size effects at topological transitions}
\label{sec:4}

Quantum topological transitions as  conventional phase transitions also exhibit finite size scaling properties~\cite{griffith}.
For a finite system close to quantum criticality, the characteristic  length $\xi$ and the finite size  $L$ are the relevant length scales.  The singular part  of its free energy $\mathcal{\delta F}_{C}(g,L)$  is expected to have a finite size contribution that  can be written as~\cite{griffith},
\begin{svgraybox}
\begin{equation}
\label{casimir scaling}
 \mathcal{\delta F}^{sing}_{C}(g,L) = \Delta_{C} L^{-(d+z-1)}f(L/\xi).
\end{equation}
\end{svgraybox}

This follows from dimensional analysis and a finite size scaling assumption. It is a natural generalization of the classical result  for the quantum case~\cite{Krech} and for topological transitions~\cite{griffith}. In Eq.~\ref{casimir scaling}, the dimension $d$ of the classical system  is replaced by the effective dimension $d+z$ as in the quantum hyperscaling relation~\cite{brankov}.
At the QCP of the bulk system, the characteristic length  is infinite, and
the scaling function $f(L/\xi=0)=1$. For  $d+z=1+1$, conformal invariance implies that the amplitudes $\Delta_C$ are universal quantities~\cite{kamenev}. 

The  scaling form, Eq.~\ref{casimir scaling},  of the finite size contribution to the free energy has been successfully verified for several systems exhibiting  topological quantum phase transitions~\cite{griffith}, as the $1d$ $p$-wave superconductor model of Kitaev,  the  $3d$ SSH model and a $3d$ model for  topological insulators~\cite{griffith}. In all these cases, the dynamic exponent  is given by $z=1$ and the correlation length exponent turns out to be $\nu=1$.

For the purpose of calculating  the finite size properties of a system, it is useful to consider it as confined within two parallel planes of area S  separated by a distance L. The  free energy  per unit area of this slab can be written as~\cite{Krech, brankov, andrea}
\begin{equation}
\label{finite size}
\lim_{S \rightarrow \infty}  \frac{ \mathcal{F} (g, L)} {J S} = L F_{bulk}(g) + F_{surface}(g) + \mathcal{\delta F}_{C}(g,L),
\end{equation}
where  $F_{bulk}(g)$ is the dimensionless bulk free energy per unit volume of the unconfined system, $F_{surface}$ is the sum of  the free energies of the surfaces,  per unit surface area, due to the confining planes and $J$ is a natural energy scale of the bulk material. If one uses periodic boundary conditions, the surface terms do not arise in the expression above~\cite{griffith}. The last term represents the finite size contribution to the free energy per unit area from the slab of width $L$. It can be imagined as mediating interface-interface interactions at a distance $L$.  Close to criticality its singular part has the scaling form given by  Eq.~\ref{casimir scaling} above. In the next sections, we use Eq.~\ref{finite size} to obtain this term and explore the physics it contains.

\subsection{Multi-band topological insulator}

In this section we focus on a $3d$ multi-band topological insulator. Recently~\cite{puel}, a theory  has been formulated that points out the existence of a time-reversal invariant in these systems with $\Theta^2=-1$.  This occurs whenever a band of conduction electrons hybridizes with the $m_J =\pm 1/2$ doublet arising from the $f$-multiplet of a rare-earth system in a crystalline environment for which this doublet is the ground state. The theory considers an effective four-band model of dispersive quasi-particles, with different effective masses. The parity of the orbitals forming these bands is such that the $k$-dependent hybridization between them is antisymmetric~\cite{puel}. The Hamiltonian belongs to class $AII$, and is characterized by a $\mathbb{Z}_2$ invariant.
The dispersion relations of the hybridized bands~\cite{puel} of the model are given by,
\begin{equation}
\label{wenergies}
\omega_{1/2}=\frac{1}{2}\left[(\epsilon_k^a+\epsilon_k^b)\pm\sqrt{(\epsilon_k^a-\epsilon_k^b)^2+4|V(k)|^2}\right],
\end{equation}
where 
$$\epsilon_k^a=-\epsilon_0^a+2t(\cos k_xa + \cos k_ya + \cos k_za),$$
$$\epsilon_k^b=\epsilon_0^b+2 \tilde{\alpha} t(\cos k_xa + \cos k_ya + \cos k_za)$$
are the dispersions of the originals non-hybridized bands.

The quantity $\tilde{\alpha}$ multiplying the hopping term above accounts for the different {\em effective masses} of the quasi-particles and $\epsilon_0^{a,b}$ are the centers of the bands.
The $k$ dependent hybridization is given by
$$|V(k)|^2= V_0^2 ( \sin^2 k_xa + \sin^2 k_ya + \sin^2 k_za),$$
where $V_0$ measures the intensity of the (antisymmetric) effective hybridization.

We consider here the simplest case of $\tilde{\alpha}=1$ and {\it inverted bands}, i.e,
$\epsilon_k^b= -\epsilon_k^a=-\epsilon_k$, such that,  $\epsilon_0^a=\epsilon_0^b=\epsilon_0$. This preserves the topological properties of the original model. In this case we get,
\begin{equation}
\omega_{1/2}=\pm\sqrt{\epsilon_k^2+|V(k)|^2}.
\end{equation}
In the continuum limit and for $k \rightarrow 0$, we obtain
$$ \epsilon_k= -\epsilon_0 + 6t -t(ak)^2=g-t (ak)^2$$
and
$$|V(k)|^2=V_0^2 (ak)^2,$$
with $k^2=k_x^2+k_y^2+k_z^2$ and $g=6t-\epsilon_0$. This model has  a topological transition at $g=0$ from a
non-trivial topological insulator for $g<0$ to a trivial one for $g>0$~\cite{puel}.
The dispersion relations close to the transition can be cast in the general form~\cite{balatsky},
\begin{equation}
\label{bala}
\omega_{1/2}/V_0 = \pm \sqrt{M^2 +(1-2MB)(ak)^2 +B^2(ak)^4},
\end{equation}
where $M=g/V_0$ and $B=t/V_0$.  Notice that at the QCP,  $M=0$ and for $k \rightarrow 0$, $\omega \propto k^z$ with the dynamic exponent $z=1$. Alternatively, at $k=0$ there is a  gap in the spectrum, $\omega \propto |g|$ that vanishes at the QCP with the {\it gap exponent} $\nu z=1$. The dispersion relations, Eq.~\ref{bala},   describe a large variety of topological insulators~\cite{balatsky}.
The ground state energy density associated with these dispersions is given by
\begin{equation}
\label{gsenergy}
f_{s}=\frac{E_{GS}}{V_0 V}= \frac{1}{(2 \pi)^3}\int d^3k \sqrt{M^2 +(1-2MB)(ak)^2 +B^2(ak)^4},
\end{equation}
where $V$ is the volume of the system.
Close to the topological transition, we introduce a characteristic length $\xi\propto M^{-1}\propto g^{-1}$, such that, the ground state energy density can be written in the scaling form,
\begin{equation}
\label{21}
f_{s}\propto \xi^{-4}\int_0^{\Lambda \xi} 4 \pi d(k \xi) (k \xi)^2 \sqrt{1 +(k\xi)^2 },
\end{equation}
where $\Lambda$ is a cut-off and we considered only the most singular terms close to the QCP. This equation can be cast in the  scaling form,
\begin{equation}
f_{s}\propto |g|^{\nu(d+z)}F[ \Lambda \xi],
\end{equation}
where  $\nu=1$, $z=1$, as identified previously and $d=3$. 

\begin{svgraybox}
Performing the integration of Eq.~\ref{21}  and taking the limit  $\Lambda \xi \rightarrow \infty$, one obtains different  contributions for the free energy,
\begin{itemize}
\item a cut-off independent term that corresponds to the scaling contribution ,  $f_S \propto |g|^{\nu(d+z)}=|g|^4$.
\item  a cut-off independent term, $f_s \propto |g|^4 \log |g|$ that violates hyperscaling~\cite{balatsky}.
\end{itemize}
Cut-off dependent contributions including,
\begin{itemize}
\item  a constant term, i.e., independent of $g$, that represents to the most singular cut-off dependent term.
\item   a term of order $|g|^2$ with a cut-off dependent coefficient. This appears for all $d \ge 1$.
\end{itemize}
\end{svgraybox}
For the one, two and three dimensional systems studied here,  the correlation length exponents take the value $\nu=1$, the dynamical exponents $z=1$ and consequently the  {\it gap exponents} $\nu z=1$. When these  are substituted in the quantum hyperscaling relation, Eq.~\ref{hyper},  we obtain that $\alpha <0$ for $d>1$. In this case the non-universal, cut-off dependent $|g|^2$ term in the free energy,  present for all $d\ge 1$,  dominates its behavior as $g \rightarrow 0$. 
Since this is the leading term for $d>1$,  then $d=1$ plays the role of an upper critical dimension for these topological transitions. According to this interpretation, we expect the critical exponents to be fixed at their $1d$ values for all $d>1$.
The presence of a logarithmic correction to the ground state energy in $d=1$ is consistent with its role as a marginal dimension.

If one considers an expansion of the more general expression  for the free energy, Eq.~\ref{gsenergy}, in powers of $B$ ($B=(t/V_0) <1$), we find that the  contribution proportional to  $|g|^4 \log |g|$ remains and acquires a $B$ dependent coefficient~\cite{balatsky}. Subtracting the diverging, cut-off dependent terms in this expansion, this simple type of renormalization  leads to a free energy $f_s(M,B)$ that exhibits a discontinuous transition between the trivial, $M<0$ and topological insulator, $M>0$ as a function of $B$~\cite{balatsky}. This  possibility of a first order topological transition associated with a gap that never closes~\cite{gapclosure} is very  interesting  and we wish to examine it using a  type of renormalization different from that of Ref.~\cite{balatsky}.

\subsection{Casimir effects in topological insulators}

The first order topological transition found in Ref.~\cite{balatsky} at $B=B_c$ relies on the renormalization procedure to deal with the cut-off in   Eq.~\ref{gsenergy}. We explore here the possibility of a discontinuous topological transition using a new scenario and a different renormalization procedure. For this purpose we consider, as in Section~\ref{sec:4}, that the system with the spectrum of excitations corresponding to Eq.~\ref{wenergies} is confined within two parallel plates of area $S$ separated by a distance $L$.
The  free energy  per unit area of this system is given by Eq.~\ref{finite size}. Here we present calculations of the quantity
$\mathcal{\delta F}^{sing}_{C}(g,L) $ for a slab of a multi-band topological insulator using a method similar to that for obtaining
the Casimir force between parallel plates in the theory of electromagnetism~\cite{abel,schi}. Since Casimir's calculation is also a renormalization procedure, we investigate the possibility of a discontinuous topological transition in the multi-band topological insulator using this approach.
\begin{figure}[b]
\sidecaption[t]
\includegraphics[scale=.8]{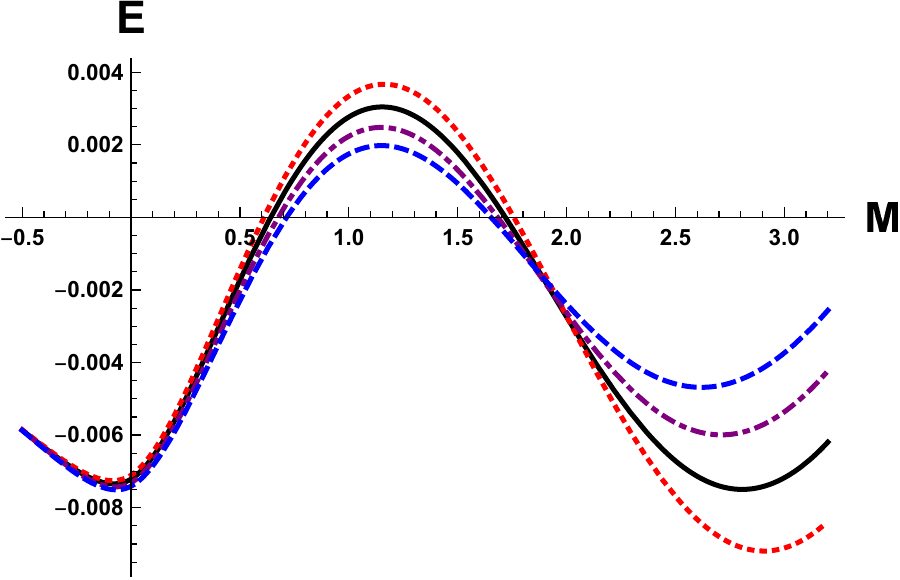}
\caption{(Color online) The ground state energy $E(M,B,L)$ as a function of $M$ for $L=3$ fixed and different values of $B$.  Dotted line (red) $B=0.41$, Full line (black) $B=B_c=0.39$, Dot-dashed line (purple) $B=0.37$ and dashed line (blue) $B=0.35$. For $B=B_c \approx 0.39$, the minimum at small, negative $M$ exchanges stability with the one at positive $M$.}
\label{fig6}       
\end{figure}
The boundary conditions in the slab are that the wave functions assume the same constant value in both planes, at $z=0$ and $z=L$.
The energy of the insulating slab can be written as,
\begin{equation}
\frac{E_S}{V_0}\!=\!\frac{4 \pi^2 S a}{L^3}\!\!\int_0^{\infty}\!\!\!dy y\!\! \sum_{n=-\infty}^{ \infty}\!\!\!
\sqrt{M_L^2+(1\!-\!2M_LB_L)(y^2+n^2)+B_L^2(y^2+n^2)^2}.
\end{equation}
where $y=k_{\perp}L / 2 \pi$, with $k_{\perp}^2=k_x^2+k_y^2$, $M_L=M(L/2 \pi a)$ and $B_L=B (2 \pi a/L)=L_0/L$ where we introduced a new length scale $L_0=2 \pi a B=2 \pi a t/V_0$ associated with the hybridization ($a$ is the lattice spacing).
The energy of the insulator occupying the whole space is given by,
\begin{equation}
\frac{E_B}{V_0}\!=\!\frac{4 \pi^2 S a}{L^3}\int_0^{\infty}\!\! dy y \int_{-\infty}^{\infty}\!\! dt \sqrt{M_L^2+(1\!-\!2M_LB_L)(y^2+t^2)+B_L^2(y^2+t^2)^2},
\end{equation}
with $t=k_zL/2 \pi$.

\begin{figure}[b]
\sidecaption[t]
\includegraphics[scale=.8]{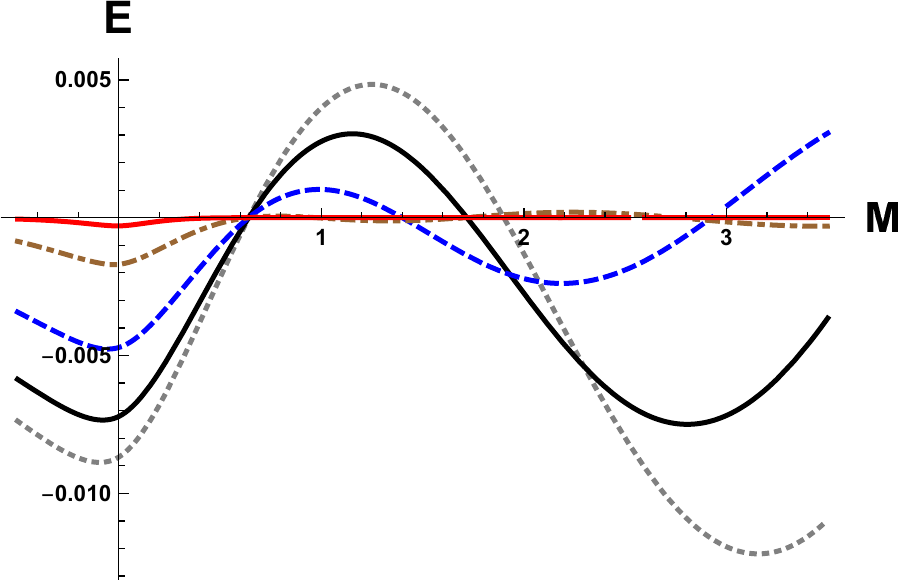}
\caption{The ground state energy $E(M,B,L)$ as a function of $M$ for different values of $L$ and $B=B_c=0.39$ fixed.  From bottom to top (at $M=0$): dotted line (grey) $L=2.8$, full line (black) $L=3$, dashed line (blue) $L=3.5$, dot-dashed line (brown) $L=5$ full line (red) $L=9$. For $L \approx 3$ there is a first order  topological transition (full line) (see text). }
\label{fig7}       
\end{figure}

The calculation of the energy difference, $\Delta E= E_S- E_B$  yields the scaling contribution according to Eq.~\ref{finite size}. It is carried out in Ref.~\cite{griffith} using the techniques to obtain the Casimir force in critical slabs. We obtain
for this energy difference at $M=0$, or $\xi=\infty$, i.e., at the topological transition
\begin{equation}
\label{finiteSS}
\frac{\Delta E}{SV_0}=\frac{- \pi^2 a}{15} L^{-3},
\end{equation}
which obeys the finite size scaling form
\begin{equation}
\label{fs1}
\frac{\Delta E}{SV_0}=\Delta_C L^{-(d+z-1)},
\end{equation}
with $d=3$, $z=1$ and the Casimir amplitude $\Delta_C = - \pi^2 a/15$. Away from criticality, since $M_L=L/\xi$, we can write
\begin{equation}
\label{fs2}
\frac{\Delta E}{SV_0}=-16 \pi^2 a L^{-3} f(L/\xi).
\end{equation}
For $L/\xi \gg 1$, the scaling function $f(L/\xi) \propto \exp(-2 \pi L/\xi)$ and the finite size contribution vanishes exponentially for $L \gg \xi$.

The full expression for the energy difference is given by~\cite{griffith},
\begin{equation}
\label{fullde}
\frac{\Delta E}{S V_0}= \frac{-32 \pi^3 B}{L^4} \left( \int_{x_2}^{x_1} dt \frac{f_1(t)}{e^{2 \pi t} - 1} + \int_{x_1}^{\infty} dt \frac{f_2(t)}{e^{2 \pi t} - 1}  \right),
\end{equation}
which is a function of $M,B$ and $L$.
The quantities $x_{1,2}$ are given by
\begin{equation}
\label{root}
x^2_{1,2}=\frac{1}{2B_L^2}\left[(1-2MB) \pm \sqrt{1-4MB} \right].
\end{equation}
The functions in the integrand are
\begin{eqnarray}
\label{f1}
f_1(t)=\frac{1}{16} x_1^4 \bigg[ \frac{\pi}{2}(1-\alpha^2)^2 - 2 \sqrt{(\eta^2-\alpha^2)(1- \eta^2)} (1-2 \eta^2+\alpha^2)- \\ \nonumber
(1- \alpha^2)^2 \tan^{-1} \frac{1-2 \eta^2 + \alpha^2 }{2 \sqrt{(\eta^2-\alpha^2)(1- \eta^2)}} \bigg],
\end{eqnarray}
for $x_2 < t <x_1$, where $\eta=t/x_1$,  $\alpha^2= (x_2^2/x_1^2)$
and
\begin{equation}
\label{f2}
f_2(t)=\frac{\pi}{16}x_1^4(1-\alpha^2)^2,
\end{equation}
that is independent of $t$ ($t >x_1$).

Finally, the  expression for the free energy difference $E(M,B,L)=\Delta E/S V_0$,  Eq.~\ref{fullde}, can be integrated numerically and the results are shown in Figs.~\ref{fig6} and ~\ref{fig7}.
In Fig.~\ref{fig6}, $E(M,B,L)$ is plotted  as a function of $M$ for a fixed separation $L=3$ between the plates and different values of the parameter $B$. One notices the presence of two minima, one at small negative values of $M$ and another for positive $M$. These minima exchange stability at a critical value of $B=B_c \approx 0.39$. The quantity $M$ plays the role of an order parameter being negative in the trivial phase and positive in the topologically non-trivial phase~\cite{balatsky}. For $B > B_c$ the stable minimum occurs for positive $M$ and the system is in the topological phase. For $B<B_c$ the minimum at small negative $M$ is the more stable and the system is in the trivial insulating phase. They exchange stability at $B=B_c$ where a first order transition occurs.

In Fig.~\ref{fig7}, $E(M,B,L)$ is plotted  as a function of $M$, now for a fixed value of $B=B_c$ and increasing separations $L$ between the plates.  For $L \approx 3$ there is a first order topological phase transition (full black line), such that, for systems  with $L  > 3$, the stable phase is the topologically trivial with $M<0$.   As $L$ increases the minimum at negative $M$ moves to zero and the curve $E(M,B_c,L \rightarrow \infty)$ has a single minimum at this value of $M$. The amplitude of the minimum at $M=0$ decreases according to the finite size scaling law, Eq.~\ref{finiteSS} and the curve for $E(M,B,L)$ becomes progressively flat and small  as a function of $M$.

A phenomenon similar to the one we have obtained, i.e.,  a first order transition in finite slabs that eventually evolves to a continuous one for large separations between plates has also been shown to occur in a strongly interacting  system~\cite{4}  exhibiting a fermionic condensate. In both cases the discontinuous character of the transition is due to finite size effects. Ultrathin films of topological insulators can provide ideal platforms to investigate these finite size effects~\cite{na3bi}.

\section{Conclusions}

In this work we discussed  how to describe and characterize topological quantum phase transitions. We  identified  a characteristic length in this problem, namely  the penetration length of the surface modes in the non-trivial topological phase of the system. It diverges as $\xi \propto |g|^{\nu}$ where $\nu$ is the correlation length exponent and $g$ the distance to the transition.  For simplicity, we neglected interactions, to put in evidence the purely topological aspects of the phenomenon   and avoid the interference of any competing long range ordering. The role of interactions in topological systems is an active area of investigation~\cite{interactions} and these may give rise to new universality classes. 

We have  obtained numerically the critical exponent $\nu=1$ for two well known systems exhibiting topological transitions, the SSH model in one dimension and the two dimensional BHZ model. Besides $\nu$,  two other critical exponents, $z$ and $\alpha$ determine the universality class of the topological transition. The former is the dynamic critical exponent that for the systems studied here  assumes the value $z=1$ implying their Lorentz invariance. This value of $z$ is also connected with the Dirac-like spectrum of excitations at the QCP. 
The exponent $\alpha$ determines the singular behavior of the free energy at zero temperature. These exponents are not independent but related through the quantum hyperscaling relation~\cite{book}. We have however pointed out that hyperscaling  can break down and indicated how this may occur for  non-interacting systems. We discussed the existence of an upper critical dimension $d_C$ for the Lorentz invariant systems treated here and argued that it takes the value $d_C=1$. We expect that for all $d >d_C$, the critical exponents  remain fixed at their values for $d=d_C$.

Finally, we have studied the possibility of discontinuous topological transitions where the gap in the spectrum never closes. Our approach is inspired on that used to study the Casimir effect, It turns out to be an efficient method of renormalization that allows to get rid of infinities. We have shown that for a $3d$ slab with one finite dimension,  finite size effects can give rise to an exchange of stability between the trivial and topological phases in a discontinuous transition. However, as the distance between the plates of the slab increases, these  effects disappear.

\begin{acknowledgement}

We would like to thank the Brazilian agencies, CNPq, CAPES and FAPERJ for partial financial support.

\end{acknowledgement}


\begin{thebibliography}{100}

\bibitem{alicea} Jason Alicea,
Rep. Prog. Phys.   {\bf 75},    076501   (2012).

\bibitem{kane} M. Z. Hasan and C. L. Kane, Rev. Mod. Phys. 82, 3045 (2010).

\bibitem{shen} Shun-Qing Shen, {\it Topological Insulators: Dirac Equation in Condensed Matter}, Second Edition, Springer Series in Solid-State Sciences, Volume 187, Springer, 2017.

\bibitem{physicab} Mucio A. Continentino, Physica B: Condensed Matter
{\bf 505},   A1-A2 (2017).

\bibitem{book} {\it Quantum Scaling in Many-Body Systems: an Approach to Quantum Phase transitions}, Mucio A. Continentino, Second Edition, Cambridge University Press, 2017.

\bibitem{koster} John Michael Kosterlitz
Rev. Mod. Phys. {\bf 89}, 040501 (? Published 9 October (2017).

\bibitem{landau}  L. D. Landau, Zh. Eksp. Teor. Fiz. {\bf 7},  19 (1937); Ukr. J. Phys. {\bf 53}, 25 (2008).

\bibitem{kitaev} A. Y. Kitaev, Physics-Uspekhi, {\bf 44}, 131 (2001); A. Kitaev, Ann. Phys., {\bf 303}, 2 (2003).

\bibitem{RG} M. A. Continentino, Fernanda Deus, Heron Caldas
Phys. Lett. {\bf A378}, 1561 (2014).

\bibitem{cristiane}  S. N. Kempkes,  A. Quelle,  C. Morais Smith,
Sci. Rep.   {\bf 6}, 38530  (2016); A. Quelle, E. Cobanera,  C. Morais Smith,
Phys. Rev. {\bf B94},  075133 (2016).

\bibitem{nandini} Mucio A. Continentino, Heron Caldas, David Nozadze, Nandini Trivedi, Physics Letters {\bf A 378},  3340 (2014).

\bibitem{chen} W. Chen, M. Legner,  A. Ruegg, and M. Sigrist, Phys. Rev. {\bf B95}, 075116  (2017).

\bibitem{chen2} Evert P. L. van Nieuwenburg, Andreas P. Schnyder, and Wei Chen, Phys. Rev. {\bf B 97}, 155151 (2018).

\bibitem{mucio} Mucio A. Continentino, G. M. Japiassu and A. Troper, Phys. Rev. {\bf 39}, 9734 (1989).

\bibitem{griffith} M. A. Griffith and M. A. Continentino, Phys. Rev. {\bf E97}, 012107 (2018).

\bibitem{analytic} Fadi Sun and Jinwu Ye, Phys. Rev. {\bf B 96}, 035113 (2017).

\bibitem{anisotropic} Bitan Roy, Pallab Goswami, and Vladimir Juri$\check{c}$i$\acute{c}$, Phys. Rev. {\bf B 95}, 201102(R) (2017); Bitan Roy and Matthew S. Foster, Phys. Rev. {\bf X 8}, 011049 (2018).

\bibitem{konig} K\"{o}nig, M., S. Wiedmann, C. Br\"{o}ne, A. Roth, H. Buhmann, L. W. Molenkamp, X.-L. Qi, and S.-C. Zhang, 2007, Science {\bf 318} (5851), 766.

\bibitem{BHZ} B. A. Bernevig, T. L. Hughes, S.-C. Zhang, Quantum Spin Hall Effect and Topological Phase Transition in HgTe Quantum Wells, Science {\bf 314} (2006), 1757.

\bibitem{bernevighughes} B. A. Bernevig with T. Hughes, Topological Insulators and Topological Superconductors. Princeton University Press, (2013);

\bibitem{hasan} M. Z. Hasan and C. L. Kane, Colloquium: Topological Insulators, Rev. Mod. Phys. {\bf 82}, 3045, (2010).

\bibitem{Imura} Ken-Ichiro Imura, Ai Yamakage, Shijun Mao, Akira Hotta, and Yoshio Kuramoto Phys. Rev. B {\bf 82}, 085118

\bibitem{Imura2} Ken-Ichiro Imura, Shijun Mao, Ai Yamakage, and Yoshio Kuramoto, Nanoscale Research Letters {\bf 6}, 358 (2011).

\bibitem{Krech} see {\it The Casimir Effect in Critical Systems}, Michael Krech, World Scientific Publishing Co. Pte. Ltd., Singapore, 1994 and references within.

\bibitem{brankov} {\it Theory of Critical Phenomena in Finite-Size
Systems: Scaling and Quantum Effects}, Jordan G. Brankov, Daniel M. Danchev, Nicholai S. Tonchev , World Scientific Publishing Co. Pte. Ltd., Singapore, 2000.

\bibitem{kamenev}  Tobias Gulden, Michael Janas, Yuting Wang, and Alex Kamenev, Phys. Rev. Lett. {\bf 116}, 026402 (2016).


\bibitem{andrea} Andrea Gambassi, J. Phys.: Conf. Ser. 161 012037 (2009).

\bibitem{puel} Griffith Mendon\c{c}a, M. A. Continentino,  and T. O. Puel, Phys. Rev. {\bf B 99}, 075109 (2019)

\bibitem{balatsky} V. Juri$\check{c}$i$\acute{c}$, D. S. L. Abergel, and A. V. Balatsky, Phys. Rev. {\bf B 95}, 161403(R) (2017).

\bibitem{gapclosure} G. Krizman, B. A. Assaf, M. Orlita, T. Phuphachong, G. Bauer, G. Springholz, G. Bastard, R. Ferreira,
L. A. de Vaulchier, and Y. Guldner, Phys. Rev.{\bf B 98}, 161202(R) (2018); G. Krizman, B. A. Assaf, T. Phuphachong, G. Bauer, G. Springholz, L. A. de Vaulchier, and Y. Guldner, Phys. Rev. {\bf B 98}, 245202 (2018).


\bibitem{abel} M. Bordag, G. L. Klimchitskaya, U. Mohideen, V. M. Mostepanenko: {\it Advances in the Casimir effect}, Oxford University Press (2009), p.22.

\bibitem{schi} J Schiefele and C Henkel, J. Phys. A: Math. Theor. {\bf 42}, 045401 (2009).

\bibitem{4} Antonino Flachi,  Muneto Nitta, Satoshi Takada, and Ryosuke Yoshii,  Phys. Rev.  Lett.  {\bf 119}, 031601 (2017); Antonino Flachi, Phys. Rev. {\bf D86}, 104047 (2012).

\bibitem{na3bi} James L. Collins, Anton Tadich, Weikang Wu, Lidia C. Gomes, Joao N. B. Rodrigues, Chang Liu, Jack Hellerstedt, Hyejin Ryu, Shujie Tang, Sung-Kwan Mo, Shaffique Adam, Shengyuan A. Yang, Michael S. Fuhrer and Mark T. Edmonds, Nature, {\bf 564}, 390 (2018).

\bibitem{interactions} Bohm-Jung Yang1, Eun-Gook Moon, Hiroki Isobe and Naoto Nagaosa1, Nat. Phys. {\bf 10} 774, (2014).


\end{thebibliography}
\end{document}